%% file: paper.tex
\documentclass[sigconf,nmanuscriptatbib=true,anonymous=false,nonacm]{acmart}

\usepackage{adjustbox}
\usepackage{enumitem}

\input{revision_commands.tex}

\input{custom_includes.tex}
\input{macros.tex}

\AtBeginDocument{%
  \providecommand\BibTeX{{%
    \normalfont B\kern-0.5em{\scshape i\kern-0.25em b}\kern-0.8em\TeX}}}

\begin{document}

\title[Blending Learning to Rank and Dense Representations for Efficient and Effective Cascades]{Blending Learning to Rank and Dense Representations\\for Efficient and Effective Cascades}

\author{Franco Maria Nardini}
\affiliation{%
  \institution{ISTI-CNR}
  \city{Pisa}
  \country{Italy}
}
\email{francomaria.nardini@isti.cnr.it}

\author{Raffaele Perego}
\affiliation{%
  \institution{ISTI-CNR}
  \city{Pisa}
  \country{Italy}
}
\email{raffaele.perego@isti.cnr.it}

\author{Nicola Tonellotto}
\affiliation{%
  \institution{University of Pisa}
  \city{Pisa}
  \country{Italy}
}
\email{nicola.tonellotto@unipi.it}

\author{Salvatore Trani}
\affiliation{%
  \institution{ISTI-CNR}
  \city{Pisa}
  \country{Italy}
}
\email{salvatore.trani@isti.cnr.it}


\input{sections/00-abstract.tex}


\maketitle

\newcommand{\ophir}[1]{\textcolor{red}{#1}}

\input{sections/01-introduction.tex}
\input{sections/03-methodology.tex}
\input{sections/04-experiments.tex}
\input{sections/05-conclusions.tex}

\section*{Acknowledgments}
This research has been partially funded by the European Union’s Horizon Europe research and innovation program EFRA (Grant Agreement Number 101093026). Views and opinions expressed are however those of the authors only and do not necessarily reflect those of the European Union or European Commission-EU. Neither the European Union nor the granting authority can be held responsible for them.

\bibliographystyle{ACM-Reference-Format}
\bibliography{sample-base}

\end{document}

%% file: revision_commands.tex
\usepackage[hide]{chato-notes}
\usepackage{soul}
\usepackage{xcolor}

%% file: custom_includes.tex
\usepackage{multirow}

%% file: macros.tex
\newcommand{\recall}{R@1000\xspace}
\newcommand{\mrr}{MRR@10\xspace}
\newcommand{\ndcg}{nDCG@10\xspace}

\newcommand{\lmart}{$\lambda$mart\xspace}

%% file: sections/00-abstract.tex

\begin{abstract}
We investigate the exploitation of both lexical and neural relevance signals for ad-hoc passage retrieval. Our exploration involves a large-scale training dataset in which dense neural representations of MS-MARCO queries and passages are complemented and integrated with $253$ hand-crafted lexical features extracted from the same corpus. 
Blending of the relevance signals from the two different groups of features is learned by a classical Learning-to-Rank (LTR) model based on a forest of decision trees.
To evaluate our solution, we employ a pipelined architecture where a dense neural retriever serves as the first stage and performs a nearest-neighbor search over the neural representations of the documents. Our LTR model acts instead as the second stage that re-ranks the set of candidates retrieved by the first stage to enhance effectiveness.
The results of reproducible experiments conducted with  state-of-the-art dense retrievers on publicly available resources show that the proposed solution significantly enhances the end-to-end ranking performance 
while relatively minimally impacting 
efficiency. Specifically, we achieve a boost in nDCG@10 of up to 11\% with an increase in average query latency of only 4.3\%. This confirms the advantage of seamlessly combining two distinct families of signals that mutually contribute to retrieval effectiveness.
\end{abstract}

%% file: sections/01-introduction.tex


\section{Introduction}
\label{sec:introduction}
Classical ranking methods rely on an inverted index containing term-level statistics such as term frequency, inverse document frequency, and positional information. These methods, known as \emph{lexical sparse retrievers}, assume vocabulary-based representations of queries and documents. Despite their effectiveness, document scoring functions based exclusively on statistical lexical signals such as BM25 fail to deal with ambiguities common in natural languages, i.e., the well-known vocabulary mismatch problem.

Previous research~\cite{10.1145/3404835.3462880,XiongXiongEtAl2021,IzacardCaronEtAl2021,HofstatterLinEtAl2021,10.1145/3572405,10.1145/3471158.3472250} demonstrates that neural ranking models based on large language models (LLMs) like BERT~\cite{DevlinEtAl2019} can accurately determine semantic similarity, and thus, substantially enhance ranking performance. These models do not explicitly model terms but estimate relevance through self-attention mechanisms by exploiting contextualized dense vector representations in low-dimensional latent spaces of query and document contents.

We explore effectively 
blending lexical and neural signals in a two-stage pipelined architecture for ad-hoc passage retrieval. Unlike standard retrieve and re-rank approaches, where the first stage performs lexical matching over an inverted index, we rely on an existing neural dense retriever to identify the candidate documents for a given query. Such candidates are then re-ranked by a re-ranking model that undergoes a training phase incorporating neural and lexical signals. The amalgamation of these signals is performed by a Learning to Rank (LTR) model based on a forest of decision trees.

\begin{figure}[htb]
    \vspace{-3mm}
  \centering
  \includegraphics[width=.9\linewidth]{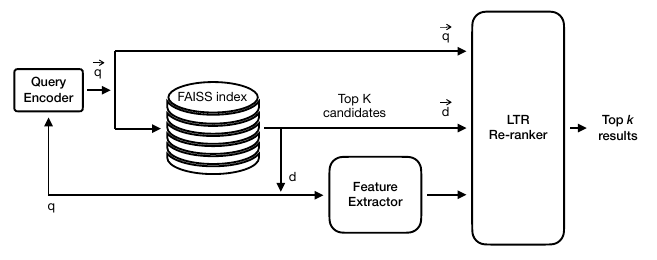} 
  \vspace{-6mm}
  \caption{Logical architecture of our system.\label{fig:architecture}}
  \vspace{-2mm}
\end{figure}

Figure \ref{fig:architecture} illustrates the organization of our end-to-end retrieval system. Given a query ${q}$, a \textit{Query Encoder} produces its dense representation $\vv{q}$. A FAISS index is then used to retrieve the $K$ document representations $\vv{d}$ most similar to $\vv{q}$. 
The representations of the query $\vv{q}$ and the $K$  documents $\vv{d}$  are then given in input to a re-ranking component, i.e., the \textit{LTR Re-ranker} in the Figure, that also employs hand-crafted features modeling the lexical matching of the query with the $K$  candidate documents. 
Such hand-crafted features are computed by the  \emph{Feature Extractor} module, also sketched in the Figure. 
The LTR re-ranker thus exploits a novel---blended---query/document representation---that encompasses neural representations of the query and the document plus the hand-crafted features modeling their lexical matching. This component is trained to combine the semantic dense representations and the lexical signals optimally. The top $k$ documents, with $k \ll K$, ordered by the re-ranking score, are finally returned as query results.

Our exploration involves a large LTR training dataset where dense neural representations of  MS-MARCO queries and passages \cite{craswell2022overview} are complemented with a set of $253$ hand-crafted lexical features already used in \cite{zhang-etal-2021-learning-rank} and blended by using a forest of decision trees.
The results of reproducible experiments, exploiting two different state-of-the-art dense representation models, i.e., STAR~\cite{10.1145/3404835.3462880} and CONTRIEVER~\cite{izacard2021contriever}, demonstrate that the proposed solution significantly enhances the end-to-end ranking performance of the neural dense system and that introducing the second-stage re-ranker does not significantly impact retrieval efficiency. 

%% file: sections/03-methodology.tex

\section{Background}
\label{sec:methodology}



Two approaches are commonly adopted to trade-off between the effectiveness of LLM-based rankers and their prohibitive
computational cost. The first approach is based on a standard pipelined architecture where an inverted index retrieves an initial set of candidate documents for each query based on a lexical scoring function such as BM25. These candidates are then re-ranked to formulate the final result list using a complex cross-attention neural ranker. The rationale of this approach is to use an efficient retriever to maximize recall and reduce the number of candidate documents to be re-ranked with the expensive---precision-oriented---neural ranker. However, cross-attention neural rankers are expensive even for re-ranking a small set of candidate documents, and devising the proper cutoff for the re-ranking stage is critical. 
The alternate approach relies on bi-encoder neural architectures, eliminating the need for the inverted index and the lexical retrieval step. In this setup, two learned encoders independently transform the content of queries and documents into dense representations within a common latent vector space~\cite{10.1145/3404835.3462880,khattab2020colbert}. Top $k$ retrieval is performed as $k$ nearest-neighbor search based on standard metrics such as inner product or cosine similarity.
Utilizing separate encoders for queries and documents enables the pre-computation of document representations, thereby shifting part of the computationally expensive processing to the indexing step. Nevertheless, such dense retrievers typically exhibit lower accuracy compared to the more expensive pipelined architectures using cross-attention neural re-rankers~\cite{INR-071}.

Although the above-mentioned LLM-based approaches exhibit state-of-the-art performance on several different retrieval tasks \cite{ThakurEtAl2021}, they 
require voluminous
labeled data for training and fail to correctly generalize term importance on out-of-domain collections or terms almost unseen during training \cite{10.1007/978-3-030-99739-7_14,ThakurEtAl2021}. The research community is thus investigating hybrid approaches encompassing the advantages of sparse lexical and dense neural retrieval models.
A prevalent approach to combining lexical and semantic ranking signals involves a straightforward linear combination of scores \cite{lin-etal-2021-batch}. 
Wang \emph{et al}. \cite{10.1145/3471158.3472233} observe that  BERT-based cross-encoders already capture the relevance signal provided by lexical models such as BM25.  Conversely, they assess the impact of interpolating BM25 and BERT-based dense retrieval scores, revealing that interpolation with BM25 is necessary for dense retrievers to perform effectively. Similarly, Askari \emph{et al}. \cite{10.1007/978-3-031-28244-7_5} find that even including the BM25 score as part of the input text enhances the re-ranking performance of BERT models.
Significantly, the authors of \cite{10.1007/978-3-030-72113-8_10} integrate the lexical retriever's score in the dense retriever's loss function. Zhang \emph{et al}.  employ lexicon-aware knowledge distillation to improve the dense encoders~\cite{10.1145/3543507.3583294}. The authors propose to do it with 1) a lexicon-augmented contrastive objective, and 2) a pair-wise rank-consistent regularization to make the dense model’s behavior incline to the lexicon one. Results on three public benchmarks show that lexicon-aware distillation strategies effectively improve the quality of the dense encoder. 
Gao \emph{et al}. follow a different strategy by proposing COIL, a novel retrieval architecture that exploits exact lexical match with query document tokens' contextualized representations~\cite{gao-etal-2021-coil}. \\
\noindent \textbf{Our contribution}. Unlike previous work, we use lexical and neural relevance
signals for ad-hoc passage retrieval differently. Specifically, we approach the problem by relying on an existing dense retrieval system to retrieve the candidate documents for a given query. We then blend dense representations with lexical sparse signals in a second re-ranking stage that employs learning-to-rank techniques, exploiting the two representations optimally.

%% file: sections/04-experiments.tex

\section{Experiments}
\label{sec:experiments}


We hypothesize that blending dense representations with hand-crafted lexical features in a classical LTR setting can improve retrieval effectiveness without hindering efficiency. We instantiate the goals of our study in the following  research questions (RQs):
\begin{itemize}
    \item[RQ1] Are LTR ranking models trained on both dense and lexical features beneficial to improve the effectiveness of a dense retrieval system in a two-stage ranking pipeline?
    \item[RQ2] Our approach relies on a second-stage re-ranker. Does this re-ranking stage impact the efficiency of end-to-end retrieval?
    \item[RQ3] Are the dense and lexical features complementary for capturing different query/passage relevance aspects?
 \end{itemize}

\begin{table*}[ht]
\caption{Effectiveness (R@1000, MRR@10, nDCG@10) and end-to-end average query latency (in msec) of the various solutions. 
Statistically significant differences (assessed with a paired $t$-test, $p < 0.01$, and Bonferroni correction) of our solutions as compared to the base models using the same number of probes are highlighted with $\dag$, while the \lmart models exploiting the full feature set as compared to the ones using only lexical or dense signals with $\S$.\label{tab:results}}
\adjustbox{max width=\textwidth}{
\begin{tabular}{lrrccccccccc}
\toprule
\multirow{2}{*}{Retriever} & \multirow{2}{*}{\# Probes} & \multirow{2}{*}{msec} & \multicolumn{3}{c}{Dev} & \multicolumn{3}{c}{DL19} & \multicolumn{3}{c}{DL20} \\
  \cmidrule(l){4-12}
& & & \recall & \mrr & \ndcg & \recall & \mrr & \ndcg & \recall & \mrr & \ndcg \\

\midrule
STAR & 20 & 11.81 & 0.8192 & 0.2830 & 0.3317 & 0.5791 & 0.8411 & 0.5432 & 0.5850 & 0.8627 & 0.5381 \\
STAR & 100 & 32.21 & 0.8822 & 0.2991 & 0.3517 & 0.6205 & 0.8760 & 0.5607 & 0.6384 & 0.8997 & 0.5651 \\
STAR & 1,000 & 257.47 & 0.9336 & 0.3120 & 0.3684 & 0.6711 & 0.9109 & 0.5793 & 0.6950 & 0.8997 & 0.5734 \\
CONTRIEVER & 1,000 & 257.23 & 0.9503 & 0.3329 & 0.3968 & 0.7448 & 0.9612 & 0.6037 & 0.7577 & 0.8966 & 
0.6215 \\

\midrule

STAR + \lmart$_{full}$ & 20 & 23.31 & 0.8192 & 0.3095$^\dag$ & 0.3627$^\dag$ & 0.5791 & 0.8837 & 0.5656 & 0.5850 & 0.8698 & 0.5527 \\
STAR + \lmart$_{full}$ & 100 & 43.71 & 0.8822 & 0.3302$^\dag$ & 0.3871$^\dag$ & 0.6205 & 0.9186 & 0.6004 & 0.6384 & 0.9138 & 0.5871 \\
STAR + \lmart$_{full}$ & 1,000 & 268.97 & 0.9336 & 0.3463$^\dag$ & 0.4089$^\dag$ & 0.6711 & 0.9651 & 0.6359 & 0.6950 & 0.9138 & 0.5925 \\

CONTRIEVER + \lmart$_{full}$ & 1,000 & 268.79 & 0.9503 & 0.3483$^\dag$ & 0.4134$^\dag$ & 0.7448 & 0.9690 & 0.6131 & 0.7577 & 0.9188 & 0.6475 \\

\midrule

STAR + \lmart$_{lexical}$ & 1,000 & 267.19 & 0.9336 & 0.3386$^\S$ & 0.4010$^\S$ & 0.6711 & 0.9535 & 0.6259 & 0.6950 & 0.9157 & 0.5908 \\
STAR + \lmart$_{dense}$ & 1,000 & 262.97 & 0.9336 & 0.3152$^\S$ & 0.3726$^\S$ & 0.6711 & 0.9070 & 0.5797 & 0.6950 & 0.9019 & 0.5828 \\

CONTRIEVER + \lmart$_{lexical}$ & 1,000 & 267.01 & 0.9503 & 0.3411$^\S$ & 0.4068$^\S$ & 0.7448 & 0.9543 & 0.6080 & 0.7577 & 0.9136 & 0.6276$^\S$ \\
CONTRIEVER + \lmart$_{dense}$ & 1,000 & 262.79 & 0.9503 & 0.3329$^\S$ & 0.3972$^\S$ & 0.7448 & 0.8814 & 0.5818 & 0.7577 & 0.9244 & 0.6297 \\

\midrule

BM25 + MonoBERT~\cite{monobert} & - & >2s (on GPU) & 0.8140 & 0.3381 & 0.3967 & 0.6778 & 0.9399 & 0.6362 & 0.6843 & 0.9259 & 0.6331 \\
BM25 + MonoELECTRA~\cite{monoelectra} & - & >2s (on GPU) & 0.8140 & 0.3474 & 0.4078 & 0.6778 & 0.9390 & 0.6317 & 0.6843 & 0.9475 & 0.6832 \\
BM25 + ELECTRA-large~\cite{electralarge} & - & >6s (on GPU) & 0.8140 & 0.3901 & 0.4483 & 0.6778 & 0.9826 & 0.6768 & 0.6843 & 0.9650 & 0.7363 \\

 \bottomrule
\end{tabular}}
\vspace{-2mm}
\end{table*}

\subsection{Dense Model and Benchmarking Datasets}
The experiments are conducted with the STAR\footnote{\url{https://github.com/jingtaozhan/DRhard}.}~\cite{10.1145/3404835.3462880} and CONTRIEVER\footnote{\url{https://huggingface.co/facebook/contriever-msmarco}.}~\cite{10.1145/3404835.3462880} dense neural models 
The models used are fine-tuned on the MS-MARCO collection for the Passage Ranking Task \cite{NguyenRosenbergEtAl2016,craswell2022overview} and encode queries and passages as single vectors in a 768-dimensional latent space. All passages in the  MS-MARCO collection were preliminarily encoded using the dense model and included in a FAISS IVF flat index ~\cite{faiss}  to implement the first-stage retriever sketched in Figure \ref{fig:architecture}. 
Such an index works by preliminarily clustering the document representations to reduce the search scope. Rather than exhaustively searching the query's nearest neighbors, the query is first compared against the centroids of the precomputed clusters. Then, only a small number of the closest clusters are probed exhaustively to determine the final result. This results in an approximate search process where the accuracy of the results and the average query latency depend on the number of clusters probed. For our experiments, we split the MS-MARCO collection into $k = 65{,}536$ clusters using the FAISS library~\cite{faiss}.

We also rely on the MS-MARCO passage-level training dataset to build the LTR training datasets. Specifically, we used the  $768$ dimensional representation as dense, semantic neural features for queries and passages and the $253$ hand-crafted features used in \cite{zhang-etal-2021-learning-rank} as sparse lexical features\footnote{Details on features and the feature extractor at \url{https://github.com/castorini/pyserini/blob/master/docs/experiments-ltr-msmarco-passage-reranking.md}.}. The resulting feature set for each query/passage pair includes in total $2{,}559$ features: $768$ features for the query and the passage representations (and the delta between their representations), the cosine similarity between the two representations, the rank of the document according to the cosine similarity, and $253$ lexical features modeling the query/passage match.
This setting provides an LTR dataset with about $500$k human-annotated queries for learning our ranking models, where dense neural representations are complemented and integrated with the hand-crafted lexical features. For each query, we included all the relevant documents (usually only 1) plus $30$ additional random negative documents from the top-1000 retrieved with  FAISS, thus obtaining a massive training dataset of about $15{.}6$M query/documents pairs. We trained the LTR models using the LightGBM\footnote{\url{https://lightgbm.readthedocs.io/en/stable/}} implementation of LambdaMART (\lmart) by optimizing the \ndcg metric on the MS-MARCO validation set. We perform hyper-parameter tuning by means of the HyperOpt\footnote{\url{http://hyperopt.github.io/hyperopt/}} library to optimize the \textit{learning rate} in [$0.01$-$0.2$], the \textit{minimal sum of the hessian in one leaf} in [$10$-$150$], and the \textit{minimum number of observations in one leaf} in [$100$-$5{,}000$]. The number of leaves is set to $64$, while the number of learned trees depends on the early stopping technique with a patience parameter set to $30$.
We assess our approach by training for each dense retriever three \lmart models that are deployed in the pipelined architecture of Figure \ref{fig:architecture}:
\begin{itemize}
    \item \lmart$_{lexical}$. This model is trained by using only the $253$ lexical features used by Zhang \emph{et al}. in \cite{zhang-etal-2021-learning-rank} plus the rank of the document provided by the first retrieval stage.
    \item \lmart$_{dense}$ This model is trained by using only neural dense features: $768$ for the query, $768$ for the passage, and $768$ for their delta. Moreover, we also include in the feature set the cosine similarity between the two representations and the rank of the document according to the cosine similarity. 
    \item \lmart$_{full}$ This model exploits the full sets of features used for either \lmart$_{lexical}$ and \lmart$_{dense}$.
\end{itemize}

As benchmarking datasets, we use three experimental collections for ad-hoc passage retrieval: TREC Deep Learning 2019 (DL19)~\cite{CraswellMitraEtAL2020}, TREC Deep Learning 2020 (DL20)~\cite{CraswellMitraEtAL2021}, and the MS-MARCO Dev set (Dev). While Dev provides a single relevant passage for each query, DL19 and DL20 provide $43$ and $54$ annotated queries, respectively, each with an average of more than $210$ passages assessed with four-grade relevance labels~\cite{craswell2022overview}. Due to the small size of DL19 and DL20, we evaluate our approach on these datasets without tuning the \lmart models to exploit the full range of graded labels and the presence of many relevant documents per query.
The latency measurements are performed on a server equipped with two Intel Xeon Silver 4314 CPU clocked at 2.40 GHz with 32 physical cores and 512 GiB of RAM. All the components of the retrieval pipeline, i.e., the query encoder, the FAISS first-stage retriever, and the feature extractor~\cite{zhang-etal-2021-learning-rank}, are executed on the CPU in a single thread. Moreover, we employ a single-threaded CPU implementation of QuickScorer~\cite{lucchese2016exploiting} that employs AVX-2 SIMD instructions to perform document scoring with \lmart models. The cross-encoder models used for comparison are instead run on an nDIVIA A100 GPU. 


\subsection{Experimental Assessment}
We experimentally answer the aforementioned research questions\footnote{The source code and the trained \lmart models will be made available upon acceptance.}.

\vspace{1mm}
\noindent \textbf{A1: improvements of ranking quality}.
To answer RQ1, Table \ref{tab:results} reports for each setting and the three benchmarking datasets the average end-to-end query latency measured in milliseconds (msec) and the retrieval performance measured in terms of R@1000, MRR@10, and nDCG@10.
The table's upper rows refer to the performance of the dense neural retriever using the STAR/CONTRIEVER representations with a FAISS IVF flat index. Specifically, for STAR (similar evidence is observed using CONTRIEVER), efficiency and effectiveness metrics are reported as a function of the number of probes (\textit{\# Probes}), i.e., the number of closest clusters visited exhaustively. As we can see from the figures in the Table, query latency is almost linearly proportional to the number of probes. At the same time, effectiveness metrics are less sensitive to the number of clusters probed.
When considering our two-stage solution, we first observe that the recall performance (R@1000) is of course the same of the base dense retriever when considering the same number of probes. 
That said, the superior performance of the proposed two-stage solution with respect to the base neural dense system emerges clearly for all settings and all datasets when considering precision-oriented metrics, i.e., nDCG@10 and MRR@10. On Dev, the second-stage re-ranker \lmart$_{full}$ constantly improves the performance by a significant margin over the neural baseline when using the same number of probes. With STAR exploiting $20$ probes, we measure a nDCG@10 increase of 9.3\%, 4.1\%, and 2.7\%, on Dev, DL19, and DL20, respectively. Such behavior is also confirmed for MRR@10. Interestingly, we observe that our entire pipeline using STAR and $20$ probes performs in nDCG@10 on Dev as the baseline STAR model with 1,000 probes but is more than one order of magnitude faster. On the contrary, with a slight increase in query latency, i.e., 4.3\%, for both the dense representational models with 1,000 probes, our solution \lmart$_{full}$ improves the retrieval effectiveness in nDCG@10 by 11\%, 9.8\%, 3.3\% using STAR, and 4.2\%, 1.6\%, 4.2\% using CONTRIEVER.

The third block of rows of the table reports the results of the ablation study conducted to understand the joint contribution of neural dense and lexical sparse features to the quality of the second-stage LTR ranker. 
Independently of the dense representation considered, the LTR models trained on the rank and lexical features only (\lmart$_{lexical}$) perform better than the models trained without lexical features (\lmart$_{dense}$). This highlights the great importance of lexical matching signals that are not entirely captured by STAR and CONTRIEVER representations. 
However, both the models learned on partial information perform worse than the model using the full set of features: their effectiveness is always lower, and the average query latency is on par.

\begin{figure}[t]
  \centering
  \includegraphics[width=.8\linewidth]{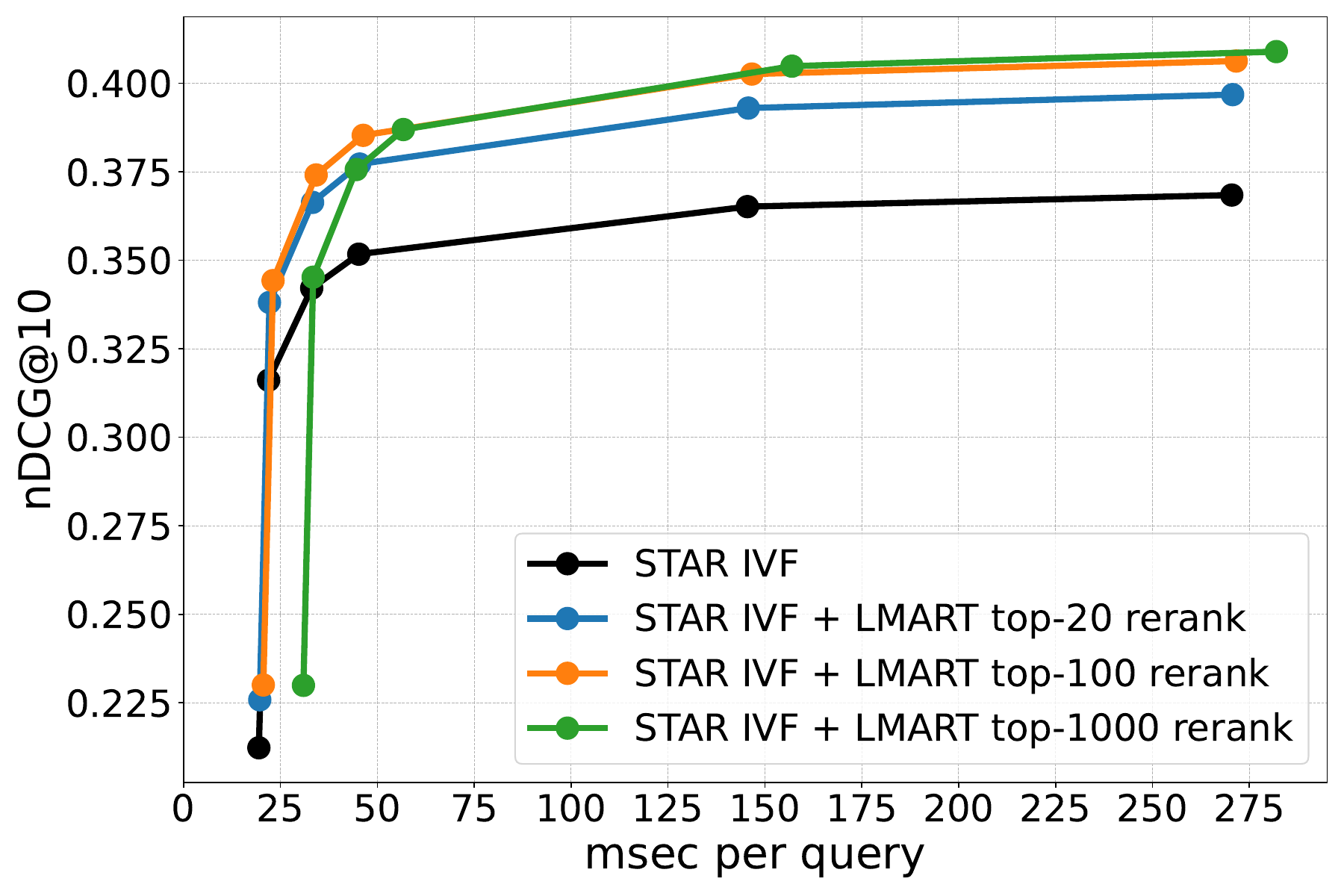}
 \vspace{-4mm}
  \caption{Efficiency/Effectiveness trade-off by varying the number of FAISS probes 
  and the  re-ranking cutoff. 
  \label{fig:overall_comparision}}
  \vspace{-6mm}
\end{figure}

\vspace{1mm}
\noindent \textbf{A2: retrieval efficiency is not impacted}.
We answer RQ2 by analyzing the end-to-end efficiency/effectiveness trade-off of our solution by varying the number of clusters probed in the FAISS index in $\{1, 10, 50, 100, 500, 1000\}$ and the number of documents re-ranked with the \lmart$_{full}$. 
Figure \ref{fig:overall_comparision} plots the efficiency-effectiveness trade-off on the Dev dataset using STAR (NDCG@10 vs. average end-to-end query latency). The behavior using CONTRIEVER is similar and not reported for brevity. We  four solid lines plotted correspond to the trade-off achieved by the first stage only (STAR IVF) and by our two-stage architecture, where the second stage re-ranks the top 20, 100, and 1,000 documents retrieved from the first stage, respectively. The six points over each solid line identify the specific performance when using 1, 10, 50, 100, 500, or 1,000 probes. The plot shows that the two-stage architecture performs best when re-ranking the top 1,000 documents. However, by observing the orange line that is shifted to the left with respect to the green one, we understand that re-ranking only the top-100 documents achieves almost identical effectiveness with a significant reduction in the average query time. As expected, a lower number of documents to re-rank positively impacts the latency of both the feature extractor and the \lmart scorer. This behavior is confirmed for all the different numbers of probes tested, i.e., orange dots are always on the left to their green counterpart. Another exciting result is observable by comparing the orange and blue lines against the black one: independently of the number of probes, our two-stage pipelines re-ranking 20 or 100 top documents outperform by a large nDCG@10 margin the FAISS baseline without any penalty in average query latency.
To complete this tradeoff analysis, the last rows of Table \ref{tab:results} report the performance of retrieval pipelines based on cross-encoder neural architectures re-ranking the top-1000 BM25 candidates. The figures reported show that our re-ranking solutions perform similarly to those based on MONO-BERT and MONO-ELECTRA. The ELECTRA-large model (24 layers, 335M params) excels in effectiveness on all three datasets instead. We note, however, that our solutions run on CPU  and are from 7$\times$ to 23$\times$  faster than those based on cross-encoders running on a high-end GPU.

\vspace{1mm}
\noindent \textbf{A3: dense and lexical features are complementary}.
To address RQ3, we present two distinct analyses exploiting the STAR representation. First, we report that among the $20$ features providing the highest gain to the \lmart$_{full}$ model, we count $13$ sparse lexical features, $6$ dense representation features, and the cosine similarity computed between the dense representations of the query and the document. Among them, the cosine is the second most important feature. This insight quantitatively highlights the complementary contribution of dense and lexical features in improving the ranking effectiveness. Indeed, although the \lmart$_{lexical}$ model exploiting only lexical features does not improve the ranking quality of the first-stage ranker, those features become 
critical when coupled with the dense ones. It is also interesting to observe that the vast majority of important dense features belong to the document representation, suggesting that neural document signals are more important than neural query signals for estimating relevance.
Second, we quantitatively analyze when \lmart$_{full}$ boosts the nDCG@10 retrieval performance on the Dev dataset the most as compared to \lmart$_{dense}$: 
86\% of the queries are not degrading their performance in terms of nDCG@10, with 25\% showing an improvement and 12\% improving the metric by at least 3 points. This analysis highlights the positive contribution of the lexical features when coupled with the dense signals.


%% file: sections/05-conclusions.tex

\section{Conclusion}
\label{sec:conclusions}
We proposed blending lexical and neural relevance signals for ad-hoc passage retrieval using Learning-to-Rank models based on forests of decision trees. We experimented with our approach by designing a novel end-to-end retrieval pipeline exploiting dense neural and sparse lexical features extracted from MS-MARCO queries and passages. Reproducible experiments show that combining the two families of signals contributes to improving retrieval effectiveness without hindering efficiency. Specifically, we achieve a boost in nDCG@10 of up to 11\% with an increase in average query latency of at most 4.3\%.

%% file: paper.bbl

\begin{thebibliography}{28}


\ifx \showCODEN    \undefined \def \showCODEN     #1{\unskip}     \fi
\ifx \showDOI      \undefined \def \showDOI       #1{#1}\fi
\ifx \showISBNx    \undefined \def \showISBNx     #1{\unskip}     \fi
\ifx \showISBNxiii \undefined \def \showISBNxiii  #1{\unskip}     \fi
\ifx \showISSN     \undefined \def \showISSN      #1{\unskip}     \fi
\ifx \showLCCN     \undefined \def \showLCCN      #1{\unskip}     \fi
\ifx \shownote     \undefined \def \shownote      #1{#1}          \fi
\ifx \showarticletitle \undefined \def \showarticletitle #1{#1}   \fi
\ifx \showURL      \undefined \def \showURL       {\relax}        \fi
\providecommand\bibfield[2]{#2}
\providecommand\bibinfo[2]{#2}
\providecommand\natexlab[1]{#1}
\providecommand\showeprint[2][]{arXiv:#2}

\bibitem[Askari et~al\mbox{.}(2023)]%
        {10.1007/978-3-031-28244-7_5}
\bibfield{author}{\bibinfo{person}{Arian Askari}, \bibinfo{person}{Amin Abolghasemi}, \bibinfo{person}{Gabriella Pasi}, \bibinfo{person}{Wessel Kraaij}, {and} \bibinfo{person}{Suzan Verberne}.} \bibinfo{year}{2023}\natexlab{}.
\newblock \showarticletitle{Injecting the BM25 Score as Text Improves BERT-Based Re-rankers}. In \bibinfo{booktitle}{\emph{Advances in Information Retrieval}}, \bibfield{editor}{\bibinfo{person}{Jaap Kamps}, \bibinfo{person}{Lorraine Goeuriot}, \bibinfo{person}{Fabio Crestani}, \bibinfo{person}{Maria Maistro}, \bibinfo{person}{Hideo Joho}, \bibinfo{person}{Brian Davis}, \bibinfo{person}{Cathal Gurrin}, \bibinfo{person}{Udo Kruschwitz}, {and} \bibinfo{person}{Annalina Caputo}} (Eds.). \bibinfo{publisher}{Springer Nature Switzerland}, \bibinfo{address}{Cham}, \bibinfo{pages}{66--83}.
\newblock
\showISBNx{978-3-031-28244-7}


\bibitem[Bruch et~al\mbox{.}(2023)]%
        {INR-071}
\bibfield{author}{\bibinfo{person}{Sebastian Bruch}, \bibinfo{person}{Claudio Lucchese}, {and} \bibinfo{person}{Franco~Maria Nardini}.} \bibinfo{year}{2023}\natexlab{}.
\newblock \showarticletitle{Efficient and Effective Tree-based and Neural Learning to Rank}.
\newblock \bibinfo{journal}{\emph{Foundations and Trends® in Information Retrieval}} \bibinfo{volume}{17}, \bibinfo{number}{1} (\bibinfo{year}{2023}), \bibinfo{pages}{1--123}.
\newblock
\showISSN{1554-0669}
\urldef\tempurl%
\url{https://doi.org/10.1561/1500000071}
\showDOI{\tempurl}


\bibitem[Clark et~al\mbox{.}(2020)]%
        {monoelectra}
\bibfield{author}{\bibinfo{person}{Kevin Clark}, \bibinfo{person}{Minh-Thang Luong}, \bibinfo{person}{Quoc~V Le}, {and} \bibinfo{person}{Christopher~D Manning}.} \bibinfo{year}{2020}\natexlab{}.
\newblock \showarticletitle{Electra: Pre-training text encoders as discriminators rather than generators}.
\newblock \bibinfo{journal}{\emph{arXiv preprint arXiv:2003.10555}} (\bibinfo{year}{2020}).
\newblock


\bibitem[Craswell et~al\mbox{.}(2021)]%
        {CraswellMitraEtAL2021}
\bibfield{author}{\bibinfo{person}{Nick Craswell}, \bibinfo{person}{Bhaskar Mitra}, \bibinfo{person}{Emine Yilmaz}, {and} \bibinfo{person}{Daniel Campos}.} \bibinfo{year}{2021}\natexlab{}.
\newblock \showarticletitle{Overview of the {TREC} 2020 deep learning track}.
\newblock \bibinfo{journal}{\emph{CoRR}}  \bibinfo{volume}{abs/2102.07662} (\bibinfo{year}{2021}).
\newblock
\showeprint[arXiv]{2102.07662}
\urldef\tempurl%
\url{https://arxiv.org/abs/2102.07662}
\showURL{%
\tempurl}


\bibitem[Craswell et~al\mbox{.}(2022)]%
        {craswell2022overview}
\bibfield{author}{\bibinfo{person}{Nick Craswell}, \bibinfo{person}{Bhaskar Mitra}, \bibinfo{person}{Emine Yilmaz}, \bibinfo{person}{Daniel Campos}, {and} \bibinfo{person}{Jimmy Lin}.} \bibinfo{year}{2022}\natexlab{}.
\newblock \showarticletitle{Overview of the TREC 2021 deep learning track}. In \bibinfo{booktitle}{\emph{Text REtrieval Conference (TREC)}}. \bibinfo{publisher}{TREC}.
\newblock
\urldef\tempurl%
\url{https://www.microsoft.com/en-us/research/publication/overview-of-the-trec-2021-deep-learning-track/}
\showURL{%
\tempurl}


\bibitem[Craswell et~al\mbox{.}(2020)]%
        {CraswellMitraEtAL2020}
\bibfield{author}{\bibinfo{person}{Nick Craswell}, \bibinfo{person}{Bhaskar Mitra}, \bibinfo{person}{Emine Yilmaz}, \bibinfo{person}{Daniel Campos}, {and} \bibinfo{person}{Ellen~M. Voorhees}.} \bibinfo{year}{2020}\natexlab{}.
\newblock \showarticletitle{Overview of the {TREC} 2019 deep learning track}.
\newblock \bibinfo{journal}{\emph{CoRR}}  \bibinfo{volume}{abs/2003.07820} (\bibinfo{year}{2020}).
\newblock
\showeprint[arXiv]{2003.07820}
\urldef\tempurl%
\url{https://arxiv.org/abs/2003.07820}
\showURL{%
\tempurl}


\bibitem[D{\'e}jean et~al\mbox{.}(2024)]%
        {electralarge}
\bibfield{author}{\bibinfo{person}{Herv{\'e} D{\'e}jean}, \bibinfo{person}{St{\'e}phane Clinchant}, {and} \bibinfo{person}{Thibault Formal}.} \bibinfo{year}{2024}\natexlab{}.
\newblock \showarticletitle{A Thorough Comparison of Cross-Encoders and LLMs for Reranking SPLADE}.
\newblock \bibinfo{journal}{\emph{arXiv preprint arXiv:2403.10407}} (\bibinfo{year}{2024}).
\newblock


\bibitem[Devlin et~al\mbox{.}(2019)]%
        {DevlinEtAl2019}
\bibfield{author}{\bibinfo{person}{Jacob Devlin}, \bibinfo{person}{Ming{-}Wei Chang}, \bibinfo{person}{Kenton Lee}, {and} \bibinfo{person}{Kristina Toutanova}.} \bibinfo{year}{2019}\natexlab{}.
\newblock \showarticletitle{{{BERT:} Pre-training of Deep Bidirectional Transformers for Language Understanding}}. In \bibinfo{booktitle}{\emph{Proc. of the 2019 Conference of the North American Chapter of the Association for Computational Linguistics: Human Language Technologies, {NAACL-HLT} 2019, Minneapolis, MN, USA, June 2-7, 2019, Volume 1 (Long and Short Papers)}}. \bibinfo{publisher}{{ACL}}, \bibinfo{pages}{4171--4186}.
\newblock
\urldef\tempurl%
\url{https://doi.org/10.18653/v1/n19-1423}
\showDOI{\tempurl}


\bibitem[Formal et~al\mbox{.}(2022)]%
        {10.1007/978-3-030-99739-7_14}
\bibfield{author}{\bibinfo{person}{Thibault Formal}, \bibinfo{person}{Benjamin Piwowarski}, {and} \bibinfo{person}{St{\'e}phane Clinchant}.} \bibinfo{year}{2022}\natexlab{}.
\newblock \showarticletitle{Match Your Words! A Study of Lexical Matching in Neural Information Retrieval}. In \bibinfo{booktitle}{\emph{Advances in Information Retrieval}}, \bibfield{editor}{\bibinfo{person}{Matthias Hagen}, \bibinfo{person}{Suzan Verberne}, \bibinfo{person}{Craig Macdonald}, \bibinfo{person}{Christin Seifert}, \bibinfo{person}{Krisztian Balog}, \bibinfo{person}{Kjetil N{\o}rv{\aa}g}, {and} \bibinfo{person}{Vinay Setty}} (Eds.). \bibinfo{publisher}{Springer International Publishing}, \bibinfo{address}{Cham}, \bibinfo{pages}{120--127}.
\newblock
\showISBNx{978-3-030-99739-7}


\bibitem[Gao et~al\mbox{.}(2021a)]%
        {gao-etal-2021-coil}
\bibfield{author}{\bibinfo{person}{Luyu Gao}, \bibinfo{person}{Zhuyun Dai}, {and} \bibinfo{person}{Jamie Callan}.} \bibinfo{year}{2021}\natexlab{a}.
\newblock \showarticletitle{{COIL}: Revisit Exact Lexical Match in Information Retrieval with Contextualized Inverted List}. In \bibinfo{booktitle}{\emph{Proceedings of the 2021 Conference of the North American Chapter of the Association for Computational Linguistics: Human Language Technologies}}, \bibfield{editor}{\bibinfo{person}{Kristina Toutanova}, \bibinfo{person}{Anna Rumshisky}, \bibinfo{person}{Luke Zettlemoyer}, \bibinfo{person}{Dilek Hakkani-Tur}, \bibinfo{person}{Iz~Beltagy}, \bibinfo{person}{Steven Bethard}, \bibinfo{person}{Ryan Cotterell}, \bibinfo{person}{Tanmoy Chakraborty}, {and} \bibinfo{person}{Yichao Zhou}} (Eds.). \bibinfo{publisher}{Association for Computational Linguistics}, \bibinfo{address}{Online}, \bibinfo{pages}{3030--3042}.
\newblock
\urldef\tempurl%
\url{https://doi.org/10.18653/v1/2021.naacl-main.241}
\showDOI{\tempurl}


\bibitem[Gao et~al\mbox{.}(2021b)]%
        {10.1007/978-3-030-72113-8_10}
\bibfield{author}{\bibinfo{person}{Luyu Gao}, \bibinfo{person}{Zhuyun Dai}, \bibinfo{person}{Tongfei Chen}, \bibinfo{person}{Zhen Fan}, \bibinfo{person}{Benjamin Van~Durme}, {and} \bibinfo{person}{Jamie Callan}.} \bibinfo{year}{2021}\natexlab{b}.
\newblock \showarticletitle{Complement Lexical Retrieval Model with Semantic Residual Embeddings}. In \bibinfo{booktitle}{\emph{Advances in Information Retrieval}}, \bibfield{editor}{\bibinfo{person}{Djoerd Hiemstra}, \bibinfo{person}{Marie-Francine Moens}, \bibinfo{person}{Josiane Mothe}, \bibinfo{person}{Raffaele Perego}, \bibinfo{person}{Martin Potthast}, {and} \bibinfo{person}{Fabrizio Sebastiani}} (Eds.). \bibinfo{publisher}{Springer International Publishing}, \bibinfo{address}{Cham}, \bibinfo{pages}{146--160}.
\newblock
\showISBNx{978-3-030-72113-8}


\bibitem[Hofst{\"{a}}tter et~al\mbox{.}(2021)]%
        {HofstatterLinEtAl2021}
\bibfield{author}{\bibinfo{person}{Sebastian Hofst{\"{a}}tter}, \bibinfo{person}{Sheng{-}Chieh Lin}, \bibinfo{person}{Jheng{-}Hong Yang}, \bibinfo{person}{Jimmy Lin}, {and} \bibinfo{person}{Allan Hanbury}.} \bibinfo{year}{2021}\natexlab{}.
\newblock \showarticletitle{Efficiently Teaching an Effective Dense Retriever with Balanced Topic Aware Sampling}. In \bibinfo{booktitle}{\emph{{SIGIR} '21: The 44th International {ACM} {SIGIR} Conference on Research and Development in Information Retrieval, Virtual Event, Canada, July 11-15, 2021}}, \bibfield{editor}{\bibinfo{person}{Fernando Diaz}, \bibinfo{person}{Chirag Shah}, \bibinfo{person}{Torsten Suel}, \bibinfo{person}{Pablo Castells}, \bibinfo{person}{Rosie Jones}, {and} \bibinfo{person}{Tetsuya Sakai}} (Eds.). \bibinfo{publisher}{{ACM}}, \bibinfo{pages}{113--122}.
\newblock
\urldef\tempurl%
\url{https://doi.org/10.1145/3404835.3462891}
\showDOI{\tempurl}


\bibitem[Izacard et~al\mbox{.}(2021a)]%
        {IzacardCaronEtAl2021}
\bibfield{author}{\bibinfo{person}{Gautier Izacard}, \bibinfo{person}{Mathilde Caron}, \bibinfo{person}{Lucas Hosseini}, \bibinfo{person}{Sebastian Riedel}, \bibinfo{person}{Piotr Bojanowski}, \bibinfo{person}{Armand Joulin}, {and} \bibinfo{person}{Edouard Grave}.} \bibinfo{year}{2021}\natexlab{a}.
\newblock \showarticletitle{Towards Unsupervised Dense Information Retrieval with Contrastive Learning}.
\newblock \bibinfo{journal}{\emph{CoRR}}  \bibinfo{volume}{abs/2112.09118} (\bibinfo{year}{2021}).
\newblock
\showeprint[arXiv]{2112.09118}
\urldef\tempurl%
\url{https://arxiv.org/abs/2112.09118}
\showURL{%
\tempurl}


\bibitem[Izacard et~al\mbox{.}(2021b)]%
        {izacard2021contriever}
\bibfield{author}{\bibinfo{person}{Gautier Izacard}, \bibinfo{person}{Mathilde Caron}, \bibinfo{person}{Lucas Hosseini}, \bibinfo{person}{Sebastian Riedel}, \bibinfo{person}{Piotr Bojanowski}, \bibinfo{person}{Armand Joulin}, {and} \bibinfo{person}{Edouard Grave}.} \bibinfo{year}{2021}\natexlab{b}.
\newblock \bibinfo{title}{Unsupervised Dense Information Retrieval with Contrastive Learning}.
\newblock
\newblock
\urldef\tempurl%
\url{https://doi.org/10.48550/ARXIV.2112.09118}
\showDOI{\tempurl}


\bibitem[Johnson et~al\mbox{.}(2019)]%
        {faiss}
\bibfield{author}{\bibinfo{person}{Jeff Johnson}, \bibinfo{person}{Matthijs Douze}, {and} \bibinfo{person}{Herv{\'e} J{\'e}gou}.} \bibinfo{year}{2019}\natexlab{}.
\newblock \showarticletitle{Billion-scale similarity search with {GPUs}}.
\newblock \bibinfo{journal}{\emph{IEEE Transactions on Big Data}} \bibinfo{volume}{7}, \bibinfo{number}{3} (\bibinfo{year}{2019}), \bibinfo{pages}{535--547}.
\newblock


\bibitem[Khattab and Zaharia(2020)]%
        {khattab2020colbert}
\bibfield{author}{\bibinfo{person}{Omar Khattab} {and} \bibinfo{person}{Matei Zaharia}.} \bibinfo{year}{2020}\natexlab{}.
\newblock \showarticletitle{Colbert: Efficient and effective passage search via contextualized late interaction over bert}. In \bibinfo{booktitle}{\emph{Proceedings of the 43rd International ACM SIGIR conference on research and development in Information Retrieval}}. \bibinfo{pages}{39--48}.
\newblock


\bibitem[Lin et~al\mbox{.}(2021)]%
        {lin-etal-2021-batch}
\bibfield{author}{\bibinfo{person}{Sheng-Chieh Lin}, \bibinfo{person}{Jheng-Hong Yang}, {and} \bibinfo{person}{Jimmy Lin}.} \bibinfo{year}{2021}\natexlab{}.
\newblock \showarticletitle{In-Batch Negatives for Knowledge Distillation with Tightly-Coupled Teachers for Dense Retrieval}. In \bibinfo{booktitle}{\emph{Proceedings of the 6th Workshop on Representation Learning for NLP (RepL4NLP-2021)}}, \bibfield{editor}{\bibinfo{person}{Anna Rogers}, \bibinfo{person}{Iacer Calixto}, \bibinfo{person}{Ivan Vuli{\'c}}, \bibinfo{person}{Naomi Saphra}, \bibinfo{person}{Nora Kassner}, \bibinfo{person}{Oana-Maria Camburu}, \bibinfo{person}{Trapit Bansal}, {and} \bibinfo{person}{Vered Shwartz}} (Eds.). \bibinfo{publisher}{Association for Computational Linguistics}, \bibinfo{address}{Online}, \bibinfo{pages}{163--173}.
\newblock
\urldef\tempurl%
\url{https://doi.org/10.18653/v1/2021.repl4nlp-1.17}
\showDOI{\tempurl}


\bibitem[Lucchese et~al\mbox{.}(2016)]%
        {lucchese2016exploiting}
\bibfield{author}{\bibinfo{person}{Claudio Lucchese}, \bibinfo{person}{Franco~Maria Nardini}, \bibinfo{person}{Salvatore Orlando}, \bibinfo{person}{Raffaele Perego}, \bibinfo{person}{Nicola Tonellotto}, {and} \bibinfo{person}{Rossano Venturini}.} \bibinfo{year}{2016}\natexlab{}.
\newblock \showarticletitle{Exploiting CPU SIMD extensions to speed-up document scoring with tree ensembles}. In \bibinfo{booktitle}{\emph{Proceedings of the 39th International ACM SIGIR conference on Research and Development in Information Retrieval}}. \bibinfo{pages}{833--836}.
\newblock


\bibitem[Nguyen et~al\mbox{.}(2016)]%
        {NguyenRosenbergEtAl2016}
\bibfield{author}{\bibinfo{person}{Tri Nguyen}, \bibinfo{person}{Mir Rosenberg}, \bibinfo{person}{Xia Song}, \bibinfo{person}{Jianfeng Gao}, \bibinfo{person}{Saurabh Tiwary}, \bibinfo{person}{Rangan Majumder}, {and} \bibinfo{person}{Li Deng}.} \bibinfo{year}{2016}\natexlab{}.
\newblock \showarticletitle{{MS} {MARCO:} {A} Human Generated MAchine Reading COmprehension Dataset}. In \bibinfo{booktitle}{\emph{Proceedings of the Workshop on Cognitive Computation: Integrating neural and symbolic approaches 2016 co-located with the 30th Annual Conference on Neural Information Processing Systems {(NIPS} 2016), Barcelona, Spain, December 9, 2016}} \emph{(\bibinfo{series}{{CEUR} Workshop Proceedings}, Vol.~\bibinfo{volume}{1773})}, \bibfield{editor}{\bibinfo{person}{Tarek~Richard Besold}, \bibinfo{person}{Antoine Bordes}, \bibinfo{person}{Artur~S. d'Avila Garcez}, {and} \bibinfo{person}{Greg Wayne}} (Eds.). \bibinfo{publisher}{CEUR-WS.org}.
\newblock
\urldef\tempurl%
\url{https://ceur-ws.org/Vol-1773/CoCoNIPS\_2016\_paper9.pdf}
\showURL{%
\tempurl}


\bibitem[Nogueira et~al\mbox{.}(2019)]%
        {monobert}
\bibfield{author}{\bibinfo{person}{Rodrigo Nogueira}, \bibinfo{person}{Wei Yang}, \bibinfo{person}{Kyunghyun Cho}, {and} \bibinfo{person}{Jimmy Lin}.} \bibinfo{year}{2019}\natexlab{}.
\newblock \showarticletitle{Multi-stage document ranking with BERT}.
\newblock \bibinfo{journal}{\emph{arXiv preprint arXiv:1910.14424}} (\bibinfo{year}{2019}).
\newblock


\bibitem[Thakur et~al\mbox{.}(2021)]%
        {ThakurEtAl2021}
\bibfield{author}{\bibinfo{person}{Nandan Thakur}, \bibinfo{person}{Nils Reimers}, \bibinfo{person}{Andreas R{\"{u}}ckl{\'{e}}}, \bibinfo{person}{Abhishek Srivastava}, {and} \bibinfo{person}{Iryna Gurevych}.} \bibinfo{year}{2021}\natexlab{}.
\newblock \showarticletitle{{{BEIR:} {A} Heterogeneous Benchmark for Zero-shot Evaluation of Information Retrieval Models}}. In \bibinfo{booktitle}{\emph{Proceedings of the Neural Information Processing Systems Track on Datasets and Benchmarks 1, NeurIPS Datasets and Benchmarks 2021, December 2021, virtual}}.
\newblock


\bibitem[Wang et~al\mbox{.}(2021b)]%
        {10.1145/3471158.3472233}
\bibfield{author}{\bibinfo{person}{Shuai Wang}, \bibinfo{person}{Shengyao Zhuang}, {and} \bibinfo{person}{Guido Zuccon}.} \bibinfo{year}{2021}\natexlab{b}.
\newblock \showarticletitle{BERT-Based Dense Retrievers Require Interpolation with BM25 for Effective Passage Retrieval}. In \bibinfo{booktitle}{\emph{Proceedings of the 2021 ACM SIGIR International Conference on Theory of Information Retrieval}} (Virtual Event, Canada) \emph{(\bibinfo{series}{ICTIR '21})}. \bibinfo{publisher}{Association for Computing Machinery}, \bibinfo{address}{New York, NY, USA}, \bibinfo{pages}{317–324}.
\newblock
\showISBNx{9781450386111}
\urldef\tempurl%
\url{https://doi.org/10.1145/3471158.3472233}
\showDOI{\tempurl}


\bibitem[Wang et~al\mbox{.}(2021a)]%
        {10.1145/3471158.3472250}
\bibfield{author}{\bibinfo{person}{Xiao Wang}, \bibinfo{person}{Craig Macdonald}, \bibinfo{person}{Nicola Tonellotto}, {and} \bibinfo{person}{Iadh Ounis}.} \bibinfo{year}{2021}\natexlab{a}.
\newblock \showarticletitle{Pseudo-Relevance Feedback for Multiple Representation Dense Retrieval}. In \bibinfo{booktitle}{\emph{Proceedings of the 2021 ACM SIGIR International Conference on Theory of Information Retrieval}} (Virtual Event, Canada) \emph{(\bibinfo{series}{ICTIR '21})}. \bibinfo{publisher}{Association for Computing Machinery}, \bibinfo{address}{New York, NY, USA}, \bibinfo{pages}{297–306}.
\newblock
\showISBNx{9781450386111}
\urldef\tempurl%
\url{https://doi.org/10.1145/3471158.3472250}
\showDOI{\tempurl}


\bibitem[Wang et~al\mbox{.}(2023)]%
        {10.1145/3572405}
\bibfield{author}{\bibinfo{person}{Xiao Wang}, \bibinfo{person}{Craig MacDonald}, \bibinfo{person}{Nicola Tonellotto}, {and} \bibinfo{person}{Iadh Ounis}.} \bibinfo{year}{2023}\natexlab{}.
\newblock \showarticletitle{ColBERT-PRF: Semantic Pseudo-Relevance Feedback for Dense Passage and Document Retrieval}.
\newblock \bibinfo{journal}{\emph{ACM Trans. Web}} \bibinfo{volume}{17}, \bibinfo{number}{1}, Article \bibinfo{articleno}{3} (\bibinfo{date}{jan} \bibinfo{year}{2023}), \bibinfo{numpages}{39}~pages.
\newblock
\showISSN{1559-1131}
\urldef\tempurl%
\url{https://doi.org/10.1145/3572405}
\showDOI{\tempurl}


\bibitem[Xiong et~al\mbox{.}(2021)]%
        {XiongXiongEtAl2021}
\bibfield{author}{\bibinfo{person}{Lee Xiong}, \bibinfo{person}{Chenyan Xiong}, \bibinfo{person}{Ye Li}, \bibinfo{person}{Kwok{-}Fung Tang}, \bibinfo{person}{Jialin Liu}, \bibinfo{person}{Paul~N. Bennett}, \bibinfo{person}{Junaid Ahmed}, {and} \bibinfo{person}{Arnold Overwijk}.} \bibinfo{year}{2021}\natexlab{}.
\newblock \showarticletitle{Approximate Nearest Neighbor Negative Contrastive Learning for Dense Text Retrieval}. In \bibinfo{booktitle}{\emph{9th International Conference on Learning Representations, {ICLR} 2021, Virtual Event, Austria, May 3-7, 2021}}. \bibinfo{publisher}{OpenReview.net}.
\newblock
\urldef\tempurl%
\url{https://openreview.net/forum?id=zeFrfgyZln}
\showURL{%
\tempurl}


\bibitem[Zhan et~al\mbox{.}(2021)]%
        {10.1145/3404835.3462880}
\bibfield{author}{\bibinfo{person}{Jingtao Zhan}, \bibinfo{person}{Jiaxin Mao}, \bibinfo{person}{Yiqun Liu}, \bibinfo{person}{Jiafeng Guo}, \bibinfo{person}{Min Zhang}, {and} \bibinfo{person}{Shaoping Ma}.} \bibinfo{year}{2021}\natexlab{}.
\newblock \showarticletitle{Optimizing Dense Retrieval Model Training with Hard Negatives}. In \bibinfo{booktitle}{\emph{Proceedings of the 44th International ACM SIGIR Conference on Research and Development in Information Retrieval}} (, Virtual Event, Canada,) \emph{(\bibinfo{series}{SIGIR '21})}. \bibinfo{publisher}{Association for Computing Machinery}, \bibinfo{address}{New York, NY, USA}, \bibinfo{pages}{1503–1512}.
\newblock
\showISBNx{9781450380379}
\urldef\tempurl%
\url{https://doi.org/10.1145/3404835.3462880}
\showDOI{\tempurl}


\bibitem[Zhang et~al\mbox{.}(2023)]%
        {10.1145/3543507.3583294}
\bibfield{author}{\bibinfo{person}{Kai Zhang}, \bibinfo{person}{Chongyang Tao}, \bibinfo{person}{Tao Shen}, \bibinfo{person}{Can Xu}, \bibinfo{person}{Xiubo Geng}, \bibinfo{person}{Binxing Jiao}, {and} \bibinfo{person}{Daxin Jiang}.} \bibinfo{year}{2023}\natexlab{}.
\newblock \showarticletitle{LED: Lexicon-Enlightened Dense Retriever for Large-Scale Retrieval}. In \bibinfo{booktitle}{\emph{Proceedings of the ACM Web Conference 2023}} (Austin, TX, USA) \emph{(\bibinfo{series}{WWW '23})}. \bibinfo{publisher}{Association for Computing Machinery}, \bibinfo{address}{New York, NY, USA}, \bibinfo{pages}{3203–3213}.
\newblock
\showISBNx{9781450394161}
\urldef\tempurl%
\url{https://doi.org/10.1145/3543507.3583294}
\showDOI{\tempurl}


\bibitem[Zhang et~al\mbox{.}(2021)]%
        {zhang-etal-2021-learning-rank}
\bibfield{author}{\bibinfo{person}{Yue Zhang}, \bibinfo{person}{ChengCheng Hu}, \bibinfo{person}{Yuqi Liu}, \bibinfo{person}{Hui Fang}, {and} \bibinfo{person}{Jimmy Lin}.} \bibinfo{year}{2021}\natexlab{}.
\newblock \showarticletitle{Learning to Rank in the Age of {M}uppets: Effectiveness{--}Efficiency Tradeoffs in Multi-Stage Ranking}. In \bibinfo{booktitle}{\emph{Proceedings of the Second Workshop on Simple and Efficient Natural Language Processing}}, \bibfield{editor}{\bibinfo{person}{Nafise~Sadat Moosavi}, \bibinfo{person}{Iryna Gurevych}, \bibinfo{person}{Angela Fan}, \bibinfo{person}{Thomas Wolf}, \bibinfo{person}{Yufang Hou}, \bibinfo{person}{Ana Marasovi{\'c}}, {and} \bibinfo{person}{Sujith Ravi}} (Eds.). \bibinfo{publisher}{Association for Computational Linguistics}, \bibinfo{address}{Virtual}, \bibinfo{pages}{64--73}.
\newblock
\urldef\tempurl%
\url{https://doi.org/10.18653/v1/2021.sustainlp-1.8}
\showDOI{\tempurl}


\end{thebibliography}
